\documentclass{article}

\usepackage{arxiv}

\usepackage[utf8]{inputenc} 
\usepackage[T1]{fontenc}    
\usepackage{hyperref}       
\usepackage{url}            
\usepackage{booktabs}       
\usepackage{amsfonts}       
\usepackage{nicefrac}       
\usepackage{microtype}      
\usepackage{lipsum}

\usepackage{amssymb}
\usepackage{graphicx}
\usepackage{balance}
\usepackage{color}
\usepackage{xcolor}
\usepackage{multirow}
\usepackage{caption}
\usepackage{textcomp}
\usepackage{url}

\usepackage{breakurl}

\usepackage{subfigure,epsfig,amsfonts,amsmath,cases,amssymb,graphics, latexsym, times, algorithmic, algorithm, bbm, color, soul,wrapfig,comment}

\title{iWash: A Smartwatch Handwashing Quality Assessment and Reminder System with Real-time Feedback in the Context of Infectious Disease}

\author{
  Sirat Samyoun, Sudipta Saha Shubha, Md Abu Sayeed Mondol, John A. Stankovic \\
  Department of Computer Science\\
  University of Virginia\\
  United States \\
  \texttt{\{ss8hf,ss7krd,mm5gg,stankovic\}@virginia.edu} \\
}

\newcommand\sys{iWash}

\begin{document}
\maketitle

\begin{abstract}
Washing hands properly and frequently is the simplest and most cost-effective interventions to prevent the spread of infectious diseases. People are often ignorant about proper handwashing in different situations and do not know if they wash hands properly. Smartwatches are found to be effective for assessing the quality of handwashing. However, the existing smartwatch based systems are not comprehensive enough in terms of achieving accuracy as well as reminding people to handwash and providing feedback to the user about the quality of handwashing. On-device processing is often required to provide real-time feedback to the user, and so it is important to develop a system that runs efficiently on low-resource devices like smartwatches. However, none of the existing systems for handwashing quality assessment are optimized for on-device processing. We present \sys, a comprehensive system for quality assessment and context-aware reminder for handwashing with real-time feedback using smartwatches. \sys~is a hybrid deep neural network based system that is optimized for on-device processing to ensure high accuracy with minimal processing time and battery usage. Additionally, it is a context-aware system that detects when the user is entering home using a Bluetooth beacon and provides reminders to wash hands. \sys~also offers touch-free interaction between the user and the smartwatch that minimizes the risk of germ transmission. We collected a real-life dataset and conducted extensive evaluations to demonstrate the performance of \sys. Compared to the existing handwashing quality assessment systems, we achieve around 12\% higher accuracy for quality assessment, as well as we reduce the processing time and battery usage by around 37\% and 10\%, respectively. 
\end{abstract}

\keywords{Handwashing, Reminder, Sensors, Smartwatch, Infectious diseases, On-device computation}

\section{Introduction}
\label{sec:intro}

Maintaining hand hygiene is crucial for preventing the spread of infectious diseases, and it can help reduce mortality, morbidity and healthcare costs \cite{allegranzi2009role}. The rise of the COVID-19 pandemic has shown how critical it is to wash hands frequently and properly to reduce the transmission of lethal viruses and germs. The pandemic has put the entire human race in an unprecedented crisis and cost many lives \cite{worldometer}. Since no vaccine has been invented yet, experts have identified frequent handwashing and social distancing as the two most effective ways to reduce the spread of the virus. Since the beginning of this pandemic, scientists have discovered even more significant facts about handwashing. A recent study shows that a country’s handwashing culture is a “very good” predictor of the magnitude of its spread of COVID-19 \cite{pogrebna2020impact}. However, adherence to proper handwashing is not as simple as it looks. According to guidelines by the World Health Organization (WHO) \cite{who-guideline-steps}, proper handwashing consists of several steps, which ensures that every area of the hands is properly disinfected. However, the average persons as well as the healthcare workers are often ignorant about proper handwashing in different situations \cite{tremblay20191187}. Moreover, since it is difficult to remember all the steps, people very often miss one or several steps, and consequently compromise their protection against viruses and germs. A recent study shows that 95\% of the people worldwide do not wash hands correctly \cite{correctwashing}, while frequent and proper handwashing could save a million deaths a year \cite{savelives}. As such, a hand hygiene assessment system that reminds users to wash hands when necessary, assesses the quality of handwashing, and provides feedback in real-time is of utmost importance, particularly in the current context of COVID-19 pandemic. 

Over the years, a number of hand hygiene assessment systems have been developed to assess the quality of handwashing events in terms of following the guidelines as well as to remind the users. Most of these systems are based on monitoring and assessing by direct observations \cite{arias2016assessment} \cite{tschudin2015compliance} \cite{szilagyi2013large}, or by a video camera placed inside the sink \cite{llorca2011vision} \cite{xia2015hand} \cite{hoey2010automated}. There are several reasons why these approaches do not work for assessing handwashing quality in the context of infectious diseases. Direct human observation is expensive and would be highly impractical in the context of social distancing. On the other hand, capturing user movements via video camera have the potential to leak sensitive information. Moreover, these approaches are not ubiquitous, have installation costs and do not work when the user is outside his home (e.g, travelling). To overcome these limitations, several smartwatch based automated systems have been developed for assessing the handwashing quality \cite{li2018wristwash} \cite{galluzzi2015hand} \cite{wang2020accurate} as well as reminding the users \cite{mondol2015harmony}. These systems utilize the sensing capabilities of the smartwatches, are ubiquitous and very much suitable for the daily life settings given the growing popularity of the consumer-grade smartwatches. 

The existing smartwatch based handwashing assessment systems have several limitations in the context of infectious disease. First, these systems focus on accuracy only, and do not provide any feedback to the user whether he/she washed hands properly or for enough duration. In different situations, such feedback could help taking immediate measures which, in turn, would be crucial to stop the further spread of the viruses and germs. For example, if a person entering home from outside does not wash hands properly, he/she could pass the germs to family and friends and infect them too. So, a real-time feedback provided to the user about handwashing could help stop further transmission of the viruses or germs. The existing systems offload the smartwatch sensor data to a cloud server and fetch the assessment results. These systems can be straightforwardly extended to provide feedback in real-time, however such systems are susceptible to network connectivity and bandwidth problems \cite{mediumai}, particularly for the smartwatches that are ubiquitous. These problems can be addressed through on-device processing that do not require network connectivity to process data and provide feedback in real-time such as our solution. 

People often forget to wash hands when necessary. For example, it is very important to wash hands after returning to home, particularly in the context of infectious disease like COVID-19 \cite{cdc-guideline-infectious}. However, existing systems lack the ability to provide reminders in such situations. We need a comprehensive system that guarantees high accuracy of handwashing quality assessment, runs efficiently on a smartwatch to facilitate real-time feedback, and reminds the user to wash hands in different contexts. In addition to that, the system has to interact with the user in a touch-free manner in order to avoid further contamination. However, there exist several challenges towards developing such a system. \looseness=-1

First, a proper handwashing involves a set of steps \cite{who-guideline-steps} where the same step can be performed differently by different people and even differently by the same person at different times. Consequently, there are confounding gestures in different steps. The diversity in the gestures for same step as well as confounding gestures in different steps make it difficult to identify which steps a user performs or misses in a handwashing event. Though this problem can be alleviated by using sensor data from both hands, wearing smartwatches on both wrists is neither convenient nor practical in free-living context. It is challenging to detect the steps of proper handwashing, particularly using a single wrist device.    \looseness=-1

Second, resources like processing capacity and battery life available in smartwatches are very low. High accuracy solutions like neural network based models usually require a significant amount of resources. It is challenging to run such models on the smartwatches with low latency in order to provide real-time feedback. 

Third, different smartwatches have different amounts of resources as well as different numbers of other applications that need to share the resources. In contrast to the solutions that require a fixed amount of resources, a solution that can trade-off between accuracy and resource requirements is more desirable because such an adaptive solution can provide room for other applications when needed as well as they can be used on devices with a wide range of resources. \looseness=-1

Fourth, a touch-free interaction between the watch and the user is needed to avoid further contamination, particularly in the context of infectious diseases. The tiny display of a smartwatch is also not suitable in providing detailed assessment results to the users. 

In this paper, we address each of the aforementioned challenges in a smartwatch based handwashing quality assessment system. We present \sys, a novel handwashing quality assessment system on a smartwatch that assesses whether the user washed hands properly in terms of WHO guidelines with high accuracy. It then provides the user with real-time feedback about the quality of the handwashing using voice, and also reminds the user to wash hands frequently and specifically in the situation when the user enters the home. The main contributions of this paper are:

\begin{itemize}
    \item \textbf{Handwashing quality assessment system on a smartwatch with high accuracy:} We present \sys, a comprehensive handwashing quality assessment and a real-time reminder system on a smartwatch. We employ a hybrid deep neural network (combination of Convolutional Neural Network (CNN) and Recurrent Neural Network (RNN)) based method to detect the handwashing quality in terms of the standard guidelines and achieve around 12\% accuracy improvement over the state-of-the-art systems.

    \item \textbf{Efficient smartwatch based system for providing immediate feedback:} For the first time in smartwatch-based handwashing quality assessment solutions, we employ model compression technique to optimize the system in terms of the smartwatch constraints to run the system faster and with lower battery usage in addition to guaranteeing high accuracy. We further show that our system is able to trade-off between accuracy and the resource requirements, and thus offers an adaptive solution as per the device specification.
    
    \item \textbf{A context-aware reminder system:} In the context of the infectious diseases, \sys~provides routine reminders for washing hands and also in the situation when the user enters home.
    
    \item \textbf{Touch-free interaction:} \sys~interacts with the users in a touch-free way, a critical requirement for a handwashing system to avoid further contamination from touching the devices. It intelligently uses voice interaction to overcome the tiny display challenge of the smartwatch and provides useful feedback to the user. 
    
    \item \textbf{Real-life data collection:} As there is no widely used publicly available dataset for handwashing quality assessment, we collected handwashing data from 14 participants in real-life setting and conducted extensive evaluations to demonstrate the performance superiority of \sys~compared to the state-of-the-art systems. 

\end{itemize}

The rest of the paper is organized as follows. Section \ref{sec:backg} provides the background knowledge to better understand the problem. Section \ref{sec:sys_overview} provides the basic overview of \sys. Section~\ref{sec:method} describes the details of each system component of \sys. Section \ref{sec:eval} experimentally evaluates the claims made in this paper. Section \ref{sec:related-works} summarizes the relevant research works. Section \ref{sec:discuss} and \ref{sec:conclusion} conclude the paper with some pointers for possible future works.

\section{Background}
\label{sec:backg}

\begin{figure}[h]
    \centering
    \includegraphics[width=0.75\columnwidth, height=0.52\textheight]{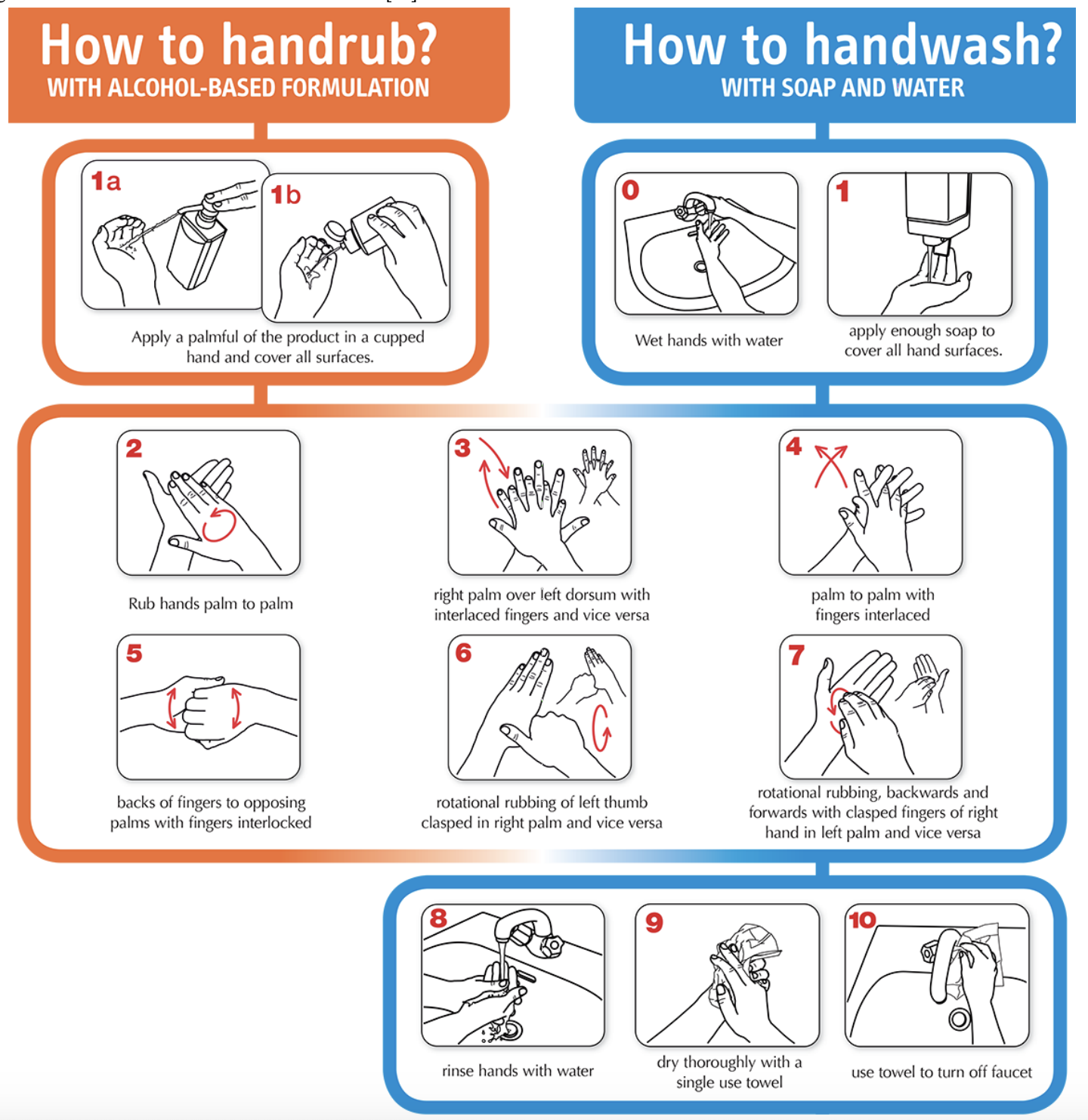}
    \caption{WHO guidelines for maintaining hand hygiene, Source: WHO. How to Handrub? / How to Handwash? ~\cite{who-guideline-steps}.}
    \label{fig:who-guideline}
\end{figure}

\begin{table}[H]
\centering
\caption{Steps for performing alcohol-based handrub and handwash with soap and water.}
\label{tab:iwashsteps}
\begin{tabular}{|l|l|l|l|}
\hline
\textbf{\begin{tabular}[c]{@{}l@{}}WHO step no.\end{tabular}} & \textbf{\begin{tabular}[c]{@{}l@{}}Alcohol-based handrub\end{tabular}} & \textbf{\begin{tabular}[c]{@{}l@{}}Handwash with soap and water\end{tabular}} & \textbf{\begin{tabular}[c]{@{}l@{}}iWash step no.\end{tabular}} \\ \hline
1                                                                 & Apply sanitizer                                                           & Apply soap                                                                      & N/A                                                            \\ \hline
2                                                                 & Rub Palm                                                                  & Rub Palm                                                                        & 1                                                              \\ \hline
3a                                                                & Rub Right Dorsum                                                          & Rub Right Dorsum                                                                & 2                                                              \\ \hline
3b                                                                & Rub Left Dorsum                                                           & Rub Left Dorsum                                                                 & 3                                                              \\ \hline
4                                                                 & Interlock fingers                                                         & Interlock fingers                                                               & 4                                                              \\ \hline
5a                                                                & Twist Right Knuckles                                                      & Twist Right Knuckles                                                            & 5                                                              \\ \hline
5b                                                                & Twist Left Knuckles                                                       & Twist Left Knuckles                                                             & 6                                                              \\ \hline
6a                                                                & Rub Right Thumb                                                           & Rub Right Thumb                                                                 & 7                                                              \\ \hline
6b                                                                & Rub Left Thumb                                                            & Rub Left Thumb                                                                  & 8                                                              \\ \hline
7a                                                                & Scrub Right Fingertip                                                     & Scrub Right Fingertip                                                           & 9                                                              \\ \hline
7b                                                                & Scrub Left Fingertip                                                      & Scrub Left Fingertip                                                            & 10                                                             \\ \hline
8                                                                 & N/A                                                                       & Rinse hands                                                                     & N/A                                                            \\ \hline
9                                                                 & N/A                                                                       & Dry hands with towel                                                            & N/A                                                            \\ \hline
10                                                                & N/A                                                                       & Turn off faucet                                                                 & N/A                                                            \\ \hline
\end{tabular}
\end{table}

\subsection{Handwashing Guidelines by WHO} 
To maintain the hand hygiene properly, WHO suggests that one should follow either handrub using an alcohol-based formulation or handwash with soap and water ~\cite{who-guideline-steps}. The alcohol-based handrub systems are the preferred mean for routine decontamination of hands for all clinical environments, while the handwash with soap and water is recommended when hands are visibly soiled~\cite{who-guideline-whywhen}. Both procedure consists of several steps which ensures that every portion of the hand is properly disinfected. The details of these steps for both approaches are shown in Figure \ref{fig:who-guideline}. While the soap or the alcohol applying step on the hands are different, we can see that the key steps for cleaning different portions of the hands (step 2 through until step 7) are the same. So, to cover both of the approaches, the existing handwashing quality assessments systems have focused on whether the steps 2 through 7 were properly performed by the user \cite{li2018wristwash} \cite{wang2020accurate}. Moreover, by carefully observing these steps, we can see that the steps 3, 5, 6, and 7 involves an action to be repeated by opposite hands. For example, the step 3 suggests that one should put the right palm over left dorsum with interlaced fingers and vice versa. It means the user should put the right palm over left dorsum with interlaced fingers and also put the left palm over right dorsum with interlaced fingers. So, for the convenience of evaluation, we break such steps into two steps and assign number for all the individual steps. The steps from the WHO handrubbing and handwashing guidelines and the corresponding step numbers in \sys~are presented in Table \ref{tab:iwashsteps}.  \looseness=-1

\subsection{What is a Quality Assessment System for Handwashing}
In most of the previous works, the adherence to the WHO handrub and handwash guidelines were measured \cite{li2018wristwash} \cite{galluzzi2015hand} \cite{wang2020accurate} \cite{mondol2015harmony} to assess the quality of hand hygiene. In other words, a handwashing or handrubbing event is assessed as a proper one if the user followed the steps in the guidelines properly. Additionally, in the context of infectious diseases, CDC(Centers for Disease Control and Prevention) recommends performing handwashing or handrubbing for a period of minimum 20 seconds \cite{cdc-guideline-infectious}. So, in this paper, by handwashing quality, we focus on the steps and the duration of the handwashing or handrubbing event performed by the user with respect to these guidelines. Moreover, for the sake of uniformity, we will use the term handwashing throughout the paper, which will cover both handwash and handrub approaches.

In a smartwatch based handwashing quality assessment system, the readings of the sensing modalities embedded in the smartwatch are used to identify the individual handwashing steps. In particular, when the user performs the handwashing steps, it involves movements of the wrists, which is captured by the IMU (Inetial Motion Unit) sensors of the smartwatch. These sensors readings are used to identify the corresponding handwashing steps.

\section{System Overview of \sys~}
\label{sec:sys_overview}

\begin{figure}[h]
    \centering
    \includegraphics[width=\columnwidth]{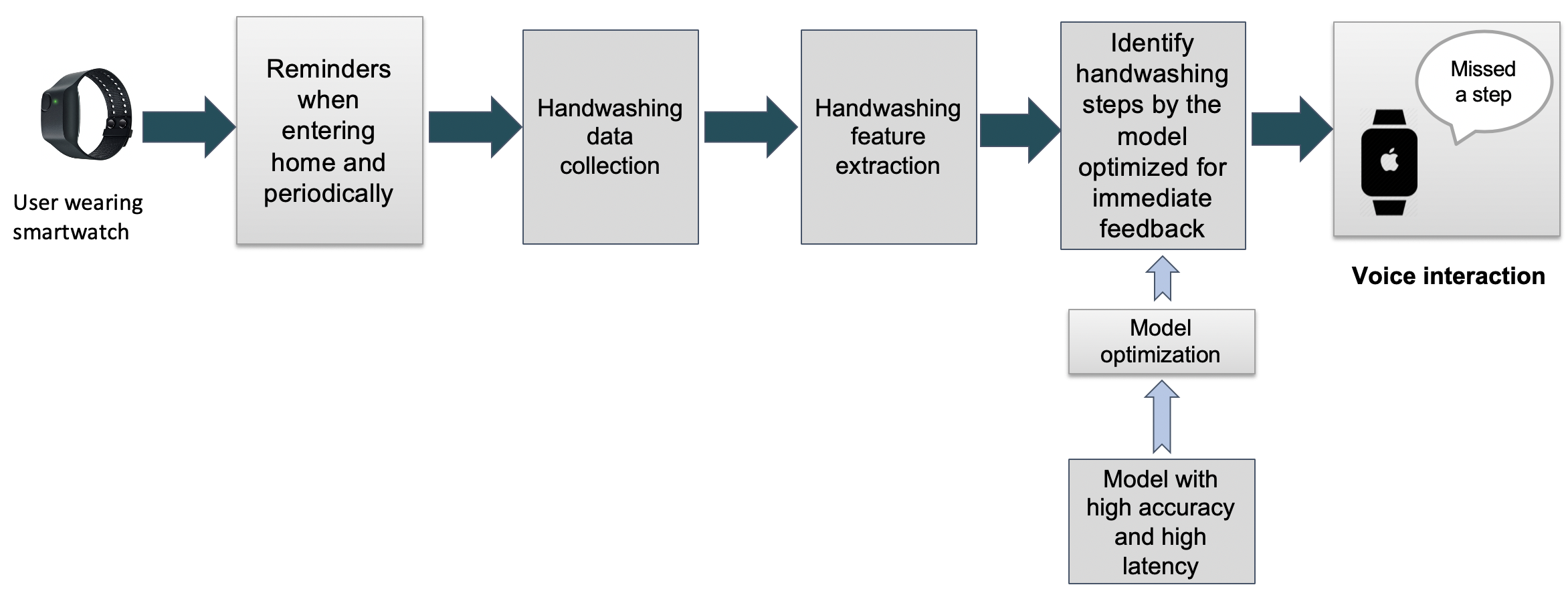}
    \caption{System overview of \sys.}
    \label{fig:sys-v}
\end{figure}

In this section we discuss the system overview of \sys. Figure \ref{fig:sys-v} depicts the overall system. The user wears a smartwatch where the \sys~application is previously installed. \sys~includes a voice interactive reminder system that is used to remind the user through the smartwatch. \sys~identifies the situation when the user enters home from outside and immediately provides a reminder to the user to wash hands. Also, the user is periodically reminded to wash hands in order to maintain the hand hygiene throughout the day. When the user starts to wash hands, \sys~collects the accelerometer, gyroscope, and magnetometer sensor readings of the smartwatch throughout the handwashing event. The start and the end of the event are marked with the help of the voice interaction with the user. Following the handwashing event, several statistical features are extracted from the data. The features are fed to a model to identify the handwashing steps following the guidelines by WHO. To build the model, we first develop a model which identifies the handwashing steps with high accuracy, however the model requires heavy computation. Then we employ a model compression technique to create an optimized version of the model for running on the low-resource smartwatch. While optimizing the model, we focus on minimizing the prediction latency to meet the requirements for providing immediate feedback as well as preserving the high accuracy to identify the handwashing steps. \sys~then identifies the steps missed by the user for a proper handwashing, and immediately reports it back to the user through a voice feedback. \sys~also reports to the user if he/she did not wash hands for a minimum duration of 20 seconds. In this way, \sys~assesses the quality of the handwashing performed by the user and provides immediate feedback to the user.

\section{System Design of \sys~}
\label{sec:method}

Based on discussion of the system overview in the previous section, the system design of \sys~has the following five major components.

\begin{itemize}
    \item Features extraction from handwashing event
    \item A hybrid deep neural network architecture for high accuracy model development
    \item A reinforcement learning based model optimization for smartwatch
    \item Identification of missing steps and feedback generation
    \item Voice interaction and reminder module  
\end{itemize}

We provide detailed description for each of the components as below:

\subsection{Features Extraction from Handwashing Event}
During a handwashing event, data are collected from the accelerometer, gyroscope and magnetometer sensors of the smartwatch. The sensor data are collected at 50Hz sampling rate. Each of the sensors provides data signals along the x, y, and z axes. Some pre-processing is needed to remove the noisy artifacts from the signals. Specifically, we pass the raw signals through a FIR (Finite Impulse Response) filter to remove the high-frequency vibration noises \cite{alam2015adaptive}. We window all the pre-processed data with a window size of 0.06s seconds, and with 70\% overlap between subsequent windows. These values are picked according to the state-of-the-art paper \cite{li2018wristwash} which led to the best results for handwashing quality assessment in their work. Once the windows is generated, we extract several statistical features from each window, such as the mean, standard deviation, kurtosis and skew. Moreover, previous works have shown that the empirical cumulative distribution function representation (ECDF) of the sensor data is effective in preserving the statistical characteristics of the data \cite{hammerla2013preserving}. Consequently, in addition to the mean, standard deviation, kurtosis and skew, we also compute the ECDF feature from the sensor data along the three axes. 


\subsection{A Hybrid Deep Neural Network Architecture for \sys} 
\label{subsec:hybrid}
\begin{figure}[h]
    \centering
    \includegraphics[width=\columnwidth]{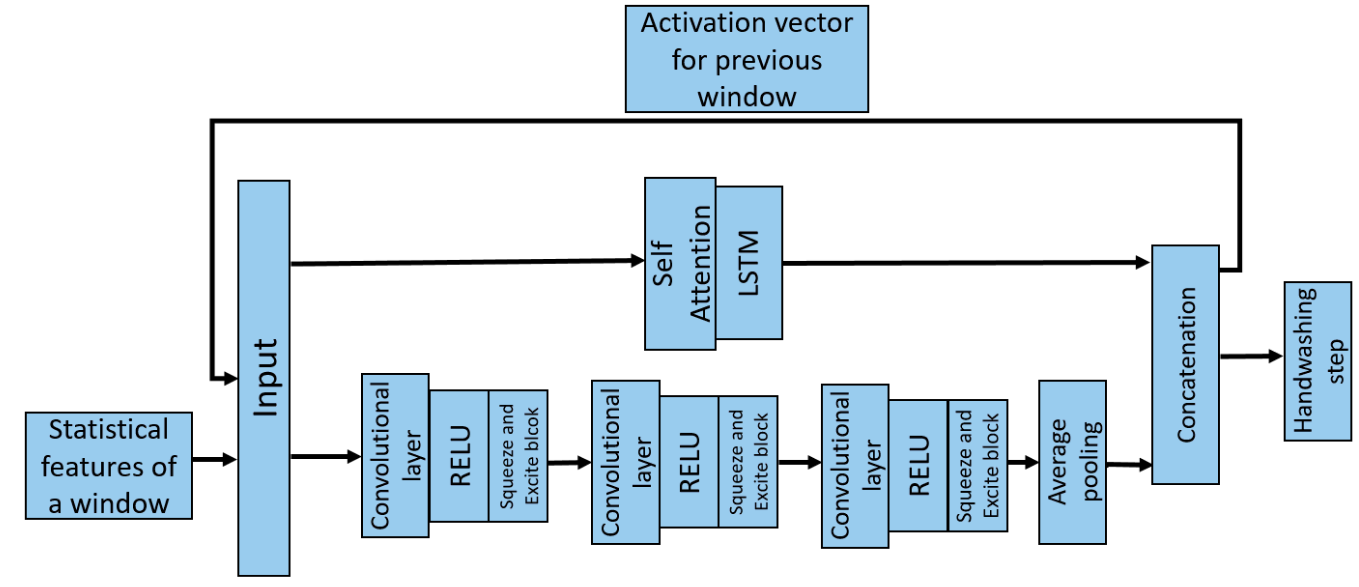}
    \caption{Hybrid deep neural network architecture for \sys.}
    \label{fig:high_acc}
\end{figure}
To develop a model that identifies handwashing steps with high accuracy, we develop a hybrid deep neural network as shown in Figure~\ref{fig:high_acc}. The model is a parallel combination of CNN and RNN. CNN and RNN allow capturing the spatial and temporal correlation present in input sensor data, respectively. As a result, the combination of these two networks facilitate the identification of handwashing steps with high accuracy.

After the sensor data are windowed from the previous step, the windowed data are provided to the hybrid deep neural network. The network identifies the handwashing step for each window. The inputs to the hybrid network are the statistical features for a window and the activation vector for the previous window data. Activation vector for a window is the concatenation of the CNN and RNN outputs for the window data. By considering the activation vector for the previous window data in addition to the statistical features of the current window data as input to the hybrid network, we allow the flow of information from the previous windows. A complete handwashing event consists of a sequence of handwashing steps, so identification of a handwashing step is facilitated if the previous handwashing steps are identified and kept into consideration. In essence, such flow of information from the previous windows facilitates the accurate identification of the subsequent steps.

\textbf{CNN Architecture:} The CNN consists of three convolutional layers. The filter size of the three convolutional layers are 3, 5, and 7, respectively. The number of filters in the three convolutional layers are 128, 256, and 512, respectively. The values of the filter size and number of filters of a convolutional layer are chosen based on evaluation on a validation set. The convolutional layers capture the correlation among data from multiple sensors across multiple axes (i.e., x, y, and z). The output of each convolutional layer passes through the Relu activation function~\cite{Agarap}. Relu activation introduces non-linearity in the network which facilitates higher accuracy. The squeeze and excite block~\cite{MALSTMFCN} after the convolutional layers help to capture contextual information present in input features. Finally, the average pooling layer reduces the number of parameters and contributes in increasing accuracy by introducing transitional invariance in the output of convolutional layer~\cite{Goodfellow-et-al-2016}.

\textbf{RNN Architecture:} We use a LSTM (Long Term Short Memory) network as the representative of RNN due to its higher performance over the other RNN variants~\cite{lstma}. The LSTM network has 50 cells, which is chosen based on evaluation on a validation set. Before the LSTM network, the input feature are passed through a self attention layer~\cite{MALSTMFCN}. The self attention layer facilitates the architecture to learn which input features need to be provided higher attention (or weights). Overall, the RNN architecture captures the temporal relationship present among input features.

\textbf{Concatenation of CNN and RNN Outputs:} The outputs of the CNN and RNN are concatenated.  As mentioned before, the concatenated vector (or activation vector) of the CNN and RNN outputs are incorporated with the input features of the next window. The concatenated vector is also passed through a fully connected network that outputs the handwashing step for the current window. The fully connected network has 3 layers, each with 250 hidden units.

\subsection{A Reinforcement Learning based Model Optimization for Smartwatch}
\label{subsec:opt}
\begin{figure}[h]
    \centering
    \includegraphics[width=0.75\columnwidth, height=0.3\textheight]{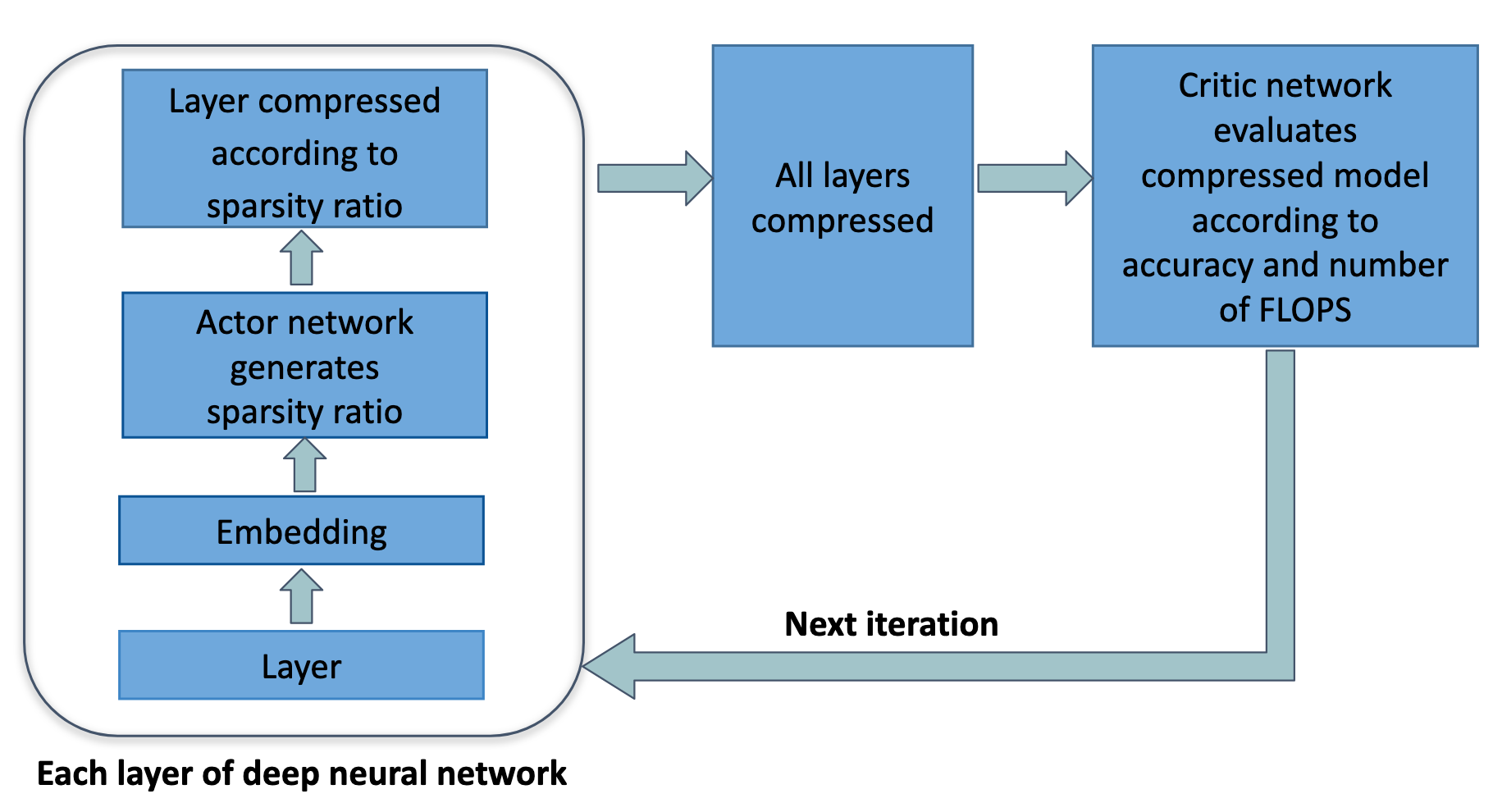}
    \caption{Overview of model optimization in \sys.}
    \label{fig:model-opt}
\end{figure}
Smartwatches are limited-resource in terms of computation and energy efficiency. To run \sys~efficiently on the smartwatch, we need to optimize the high accuracy model described in the previous section so that it runs with minimal prediction latency and battery resource. We also need to preserve the accuracy of the high accuracy model during this optimization. To overcome this challenge, we employ a model compression technique that compresses the high accuracy model. \looseness=-1

Specifically, we adopt AMC~\cite{he2018amc} as this is a state-of-the-art work on model compression focusing on the optimization of both accuracy and prediction latency. As battery usage depends on the computation duration, optimizing for prediction latency also ensures optimization for battery usage. 

In a deep learning model, the number of  FLoating point OPerations (FLOPs) represents how many computations (multiplications and additions) are required for the model. 
Thus, the greater the number of FLOPs, the higher the prediction latency is. We need to set the maximum number of FLOPs allowed in the optimized model. Specifically, we consider a hyper-parameter $\alpha\in(0, 1)$ that represents the fraction of FLOPs of the high accuracy model allowed in the optimized model. Thus, if the number of FLOPs of the high accuracy model is $N$ and we set $\alpha=0.5$, then the maximum number of FLOPs allowed in the optimized model is $0.5*N$.


We leverage reinforcement learning to efficiently and automatically explore the large design space of deep neural network models with the goal of finding a model that is optimal in terms of both accuracy and prediction latency. 
Figure~\ref{fig:model-opt} provides an overview of the reinforcement learning based model optimization procedure.
At each iteration of the reinforcement learning, we compress the high accuracy model in a layer-by-layer manner by following an actor-critic algorithm. For each layer, we generate an embedding of the layer. The embedding depends on the structure of the layer, number of FLOPs of the layer, total number of FLOPs reduced in previous layers, and the number of remaining FLOPs in the subsequent layers.
Based on the embedding, the actor network generates a sparsity ratio (a real value between 0 and 1) for the layer. The layer is compressed with the sparsity ratio. The higher the sparsity ratio, the more the layer is compressed. If the sparsity ratio is $s$, the layer will be compressed by $s*100\%$. 
Then the next layer is processed in the similar manner. After all the layer are compressed, the critic network evaluates the resultant model in terms of accuracy and number of FLOPs. During the evaluation, it is also checked whether the number of FLOPs is less than the maximum allowed. 
We run 100 such iterations. Finally, the reinforcement learning agent learns the optimal sparsity ratio for each layer of the high accuracy model. Each layer is compressed based on the corresponding sparsity ratio to generate the final optimized model.

Specifically, at each iteration of the reinforcement learning, we try to minimize the following loss function:
\begin{equation}
\label{eq:main_loss}
\mathcal{L} = L(\mathbf{y},O_C(\mathbf{x}|\mathbf{s}))*\log(\#FLOPs)
\end{equation}
where,
\begin{equation}
\label{eq:sparsity}
    \mathbf{s} = O_A(\mathbf{E})
\end{equation}
In Equation~\eqref{eq:main_loss}, $\mathbf{y}$ and $\mathbf{x}$ represent the ground truth and input features, respectively. $\mathbf{s}$ denotes the sparsity ratio generated by actor network ($O_A$, Equation~\eqref{eq:sparsity}) given embedding of the model layers ($\mathbf{E}$, Equation~\eqref{eq:sparsity}). $L$ denotes the mean squared loss between ground truth ($\mathbf{y}$) and the output generated by critic network ($O_C$). $\#FLOP$ represents the number of FLOPs in the model after compressing all the layers according to the corresponding sparsity ratios. Each of the actor and critic networks has two hidden layers, each layer has 300 units.

Overall, in Equation~\eqref{eq:main_loss}, $\mathcal{L}$ is the multiplication of two terms. The first term finds the loss between ground truth $\mathbf{y}$ and the output generated by critic network $O_C$ where the critic network considers the action (sparsity ratio for each layer, $\mathbf{s}$) generated by the actor network. The second term is the logarithm of the number of FLOPs in the model after compressing each layer according to the corresponding sparsity ratio. Thus if we minimize Equation~\eqref{eq:main_loss}, we try to maximize accuracy and minimize number of FLOPs.
 We use Adam optimizer~\cite{kingma2014adam} to minimize Equation~\eqref{eq:main_loss}.
By the minimization of Equation~\eqref{eq:main_loss}, we find the optimal parameter values for the actor and critic networks. When both the actor and critic networks have optimal parameter values, the action (sparsity ratio for each layer) generated by the actor network is also optimal. The optimal sparsity ratios ensure the optimal compressed model.


As mentioned before, choosing the value of $\alpha$ is a design choice. With the increase of the value of this hyper-parameter, the prediction latency increases but the accuracy also increases as the model is allowed to be more complex. Due to such trade-off between accuracy and number of FLOPs which is controllable by tuning the value of $\alpha$, \sys~has the capability of being device adaptive. For low-resource devices, we can choose a lower value of $\alpha$ resulting in reduced number of FLOPs. This ensures that the model will run efficiently on the device with a little sacrifice of accuracy. On the other hand, for the devices having high available resources, setting a higher value of $\alpha$ will involve higher number of FLOPs and will result in high accuracy. The details of this trade-off and device adaptivity are discussed in Section~\ref{subsubsec:adaptive} and Figure~\ref{fig:device_adaptive} illustrates the trade-off.
For our scenario, we consider $\alpha=$ 0.5 as it provides a satisfactory trade-off between accuracy and the number of FLOPs.

\subsection{Identification of Missing Steps and Feedback Generation}

\begin{table}[h]
\centering
\caption{Feedback provided to user for the corresponding quality assessed by \sys.}
\begin{tabular}{|l|l|}
\hline
\textbf{\begin{tabular}[c]{@{}l@{}}Assessed\\ quality\end{tabular}}                     & \textbf{Feedback provided to user}                              \\ \hline
Missed step 1                                                                           & Didn't rub both hands palm to palm                              \\ \hline
Missed step 2                                                                           & Didn't rub right palm over left dorsum properly                 \\ \hline
Missed step 3                                                                           & Didn't rub left palm over right dorsum properly                 \\ \hline
Missed step 4                                                                           & Didn't put palm to palm with fingers interlaced properly        \\ \hline
Missed step 5                                                                           & Didn't clean right fingertips interlocked in left palm properly \\ \hline
Missed step 6                                                                           & Didn't clean left fingertips interlocked in right palm properly \\ \hline
Missed step 7                                                                           & Didn't rub left thumb clasped in right palm properly            \\ \hline
Missed step 8                                                                           & Didn't rub right thumb clasped in left palm properly            \\ \hline
Missed step 9                                                                           & Didn't rotationally rub right fingers on left palm properly     \\ \hline
Missed step 10                                                                          & Didn't rotationally rub left fingers on right palm properly     \\ \hline
\begin{tabular}[c]{@{}l@{}}Duration less than\\ 20 seconds\end{tabular}                 & Didn't wash hands for enough duration                           \\ \hline
\begin{tabular}[c]{@{}l@{}}No missed steps\\ and duration at\\ least 20 seconds\end{tabular} & Great job! You washed your hands perfectly.                     \\ \hline
\end{tabular}
\label{tab:feedback-msg}
\end{table}

As per the claim of this paper, \sys~assesses the quality of handwashing in terms of the steps in the WHO guidelines and the duration of the handwashing. Using the model described above, \sys~identifies the steps from a handwashing event performed by a user on the smartwatch. It then compares these steps to those in the WHO guidelines, and thus finds which step(s) has been missed by the user, if any. It then generates a feedback message for each of the missing step which contains a brief description about that step. Moreover, \sys~calculates the duration of the handwashing event from the start and the end of the handwashing found from the voice interaction with the user. If the duration is less than 20 seconds, a feedback message is generated mentioning that the user did not wash hands for enough duration. However, if there are no missing steps and the duration is at least 20 seconds, a feedback message is generated mentioning that the user washed hands perfectly. Table \ref{tab:feedback-msg} shows the list of the feedback messages provided by \sys~to the user. If there are multiple feedback messages, the messages are combined into a single message at the end. The feedback is then provided to the user using voice at the end of the interaction with the user.

\subsection{Reminder and Interaction Module}
\begin{figure}[h]
    \centering
    \includegraphics[width=0.7\columnwidth]{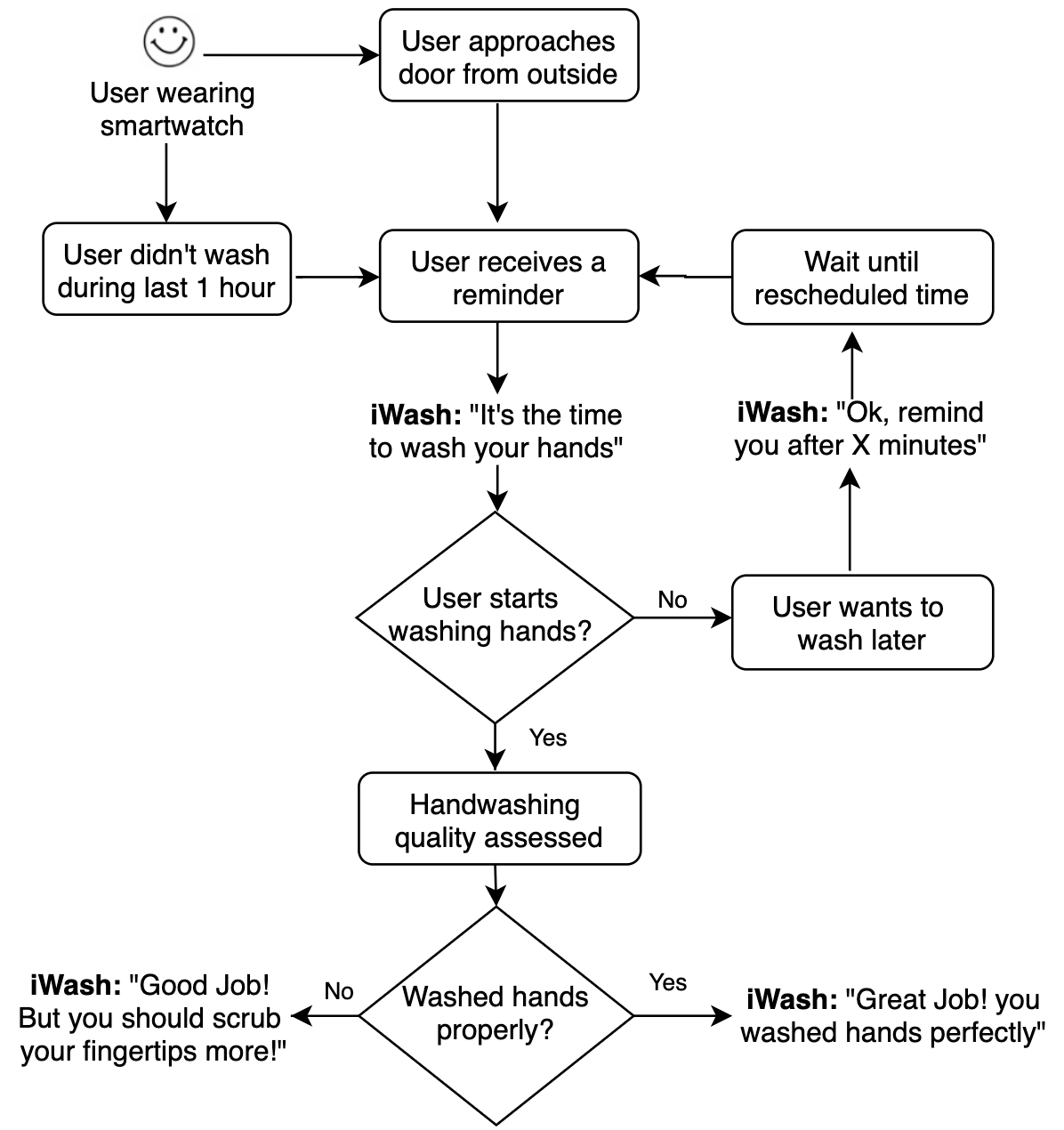}
    \caption{Flowchart of the reminder and interaction module in \sys.}
    \label{fig:flowchart_reminder}
\end{figure}

This module of \sys~is used to remind the user for washing hands in different situations and interacting with the user throughout the handwashing event. The flow of a reminder in \sys~is demonstrated in Figure \ref{fig:flowchart_reminder}. When the user wearing the watch approaches the door from the outside, he/she receives a reminder on the watch. The user is also reminded to wash hands if \sys~haven't recorded any handwashing event during the last one hour. If the user confirms starting handwashing during the interaction, \sys~collects the data during the handwashing event and assesses the quality of the handwashing, which is immediately provided the to user in the form of a voice feedback. Moreover, during the interaction, if the user wants to wash hands later at a specific time, \sys~reminds the user accordingly at the rescheduled time. The detailed functionalities of the reminder and interaction module is discussed below. \looseness=-1

\subsubsection{Providing Reminders when the User is Entering Home}
\sys~includes a low-energy, proximity Bluetooth beacon~\cite{estimote} fitted at the door of a room. The proximity beacon is a small wireless sensor that can be attached to any object. It broadcasts tiny radio signals which can be sensed by a smartwatch. Each sensed beacon signal on the smartwatch is associated with a Received Signal Strength (RSSI) value, which indicates how distant the beacon is from the smartwatch. The beacon has a unique advertisement identifier, which is programmed into the smartwatch and is used to identify the specific beacon. The \sys~application installed on the smartwatch runs in background which continuously receives signals from any surrounding beacon. Therefore, when a user wearing the smartwatch approaches the door, \sys~receives strong signals from the door beacon continuously. We use a simple scheme for detecting the home entrance. 
If the average RSSI value received from a beacon within the last 15 seconds is greater than a threshold value (-60 decibels), and if the beacon identifier matches with that of the door beacon, \sys~identifies the situation as the user is entering the home. Here the parameters, such as, -60dB and 15 seconds are picked as a design choice, which led to satisfying results for the detection of the home entrance situation compared to other situations, such as the home exit. Moreover, on detecting the home entry of the user, \sys~invokes the reminder system to remind him/her for washing hands.

\subsubsection{Providing Routine Reminders}
In the context of the infectious disease, it is also important to wash hands frequently throughout the day \cite{cdc-guideline-infectious}, whether the user is inside or outside the home. Moreover, smartwatches are very effective for notifying the user in all situation. \sys~provides daily handwashing reminders to the user between 9 AM to 9 PM. It picks the reminder times intelligently. Whenever the user performs a handwashing event following a voice interaction, \sys~records the time of the event as the last handwashing time. One hour from that time, it reminds the user with a voice message that it would be a good idea to wash hands now, since he/she did not wash hands during the last hour. This design choice ensures that the user is reminded in a routine way, but he/she is not bothered with too many reminders.

\begin{figure}[h]
    \centering
    \includegraphics[width=0.55\columnwidth]{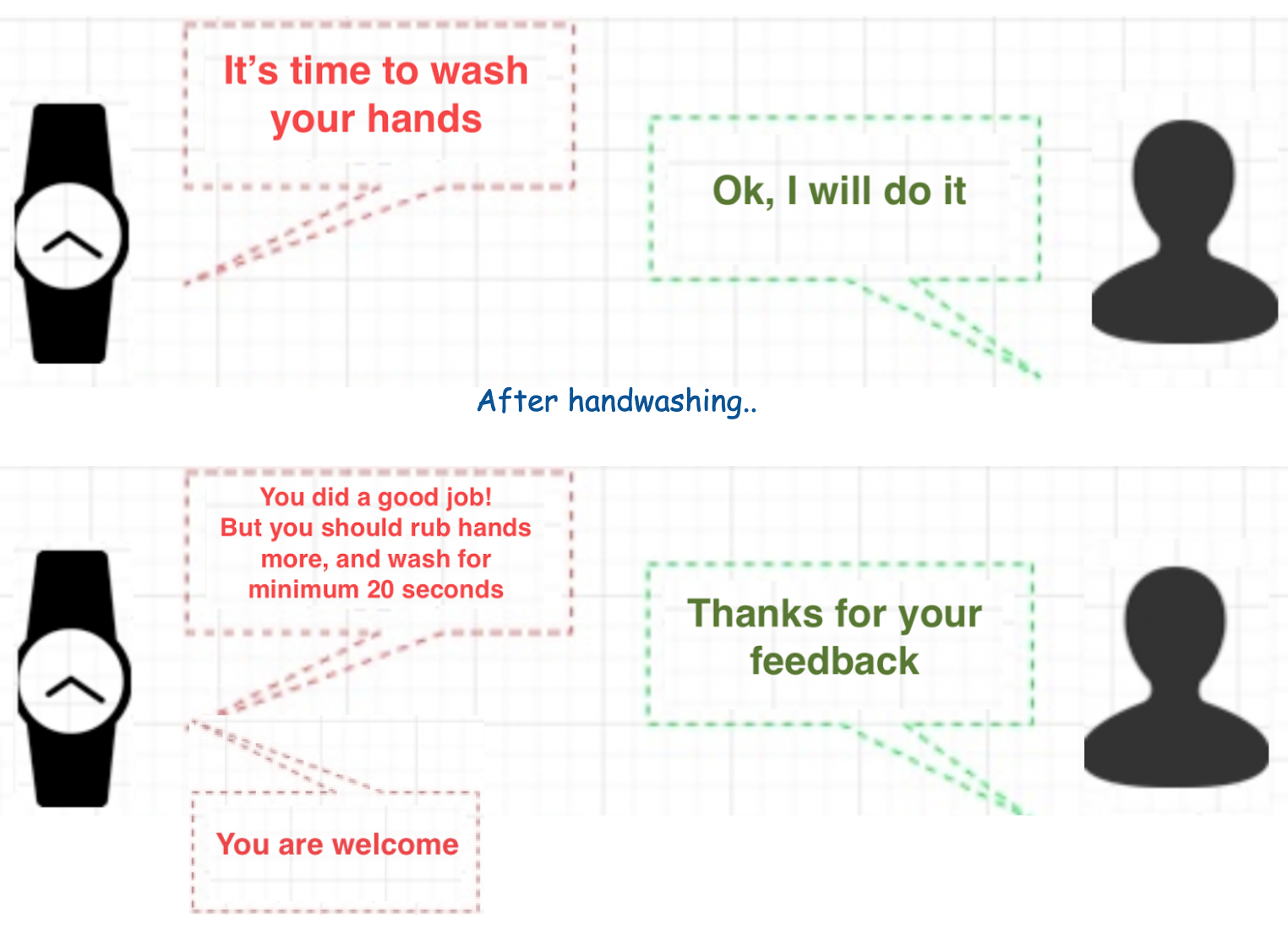}
    \caption{Flow of interaction of \sys~ with the user.}
    \label{fig:example-interaction}
\end{figure}

\subsubsection{Voice Interaction with the User}
The small display size of the smartwatch display poses a serious challenge to interact with the user and to provide the detailed feedback about handwashing. In order to overcome this challenge, \sys~uses voice input and output instead of any text for interaction. The voice interaction also ensures that the user is interacted in a touchless manner, which is critical to avoid further contamination of any virus or germ. The voice interaction also ensures that the user is interacted in a touch-free way, which is critical to avoid further contamination of any germs or virus. While providing the reminder, \sys~provides a local notification with a haptic vibration to alert the user. It then initiates the voice interaction with the user and asks him/her to wash hands. It uses the speaker of the smartwatch and leverages a text-to-speech framework for this purpose. When the user responds using any voice utterance, \sys~takes the input via the microphone of the smartwatch, and converts the audio to text using a speech-to-text framework. Then it parses the text using a keyword based mechanism similar to \cite{mondol2016medrem} \cite{samyoun2019iadhere} to understand the intent of the user. When the user confirms starting the handwashing, \sys~starts collecting data from the accelerometer, gyroscope, and magnetometer sensors of the smartwatch. When the user confirms being done with handwashing, it stops collecting data and begins the quality assessment with the collected data. An example of the interaction between \sys~and the user is demonstrated in Figure \ref{fig:example-interaction}. \looseness=-1

\subsubsection{Postponing a Reminder}
The user may not be in a situation or may not be willing to wash hands immediately after being reminded. In such cases, \sys~allows the user to snooze a reminder at a later time using voice. The user can do so by simply asking \sys~to remind after a specific number of minutes. \sys~finds out the specific time the user want to be reminded from the utterance text and sets the reminder accordingly. This design choice ensures that the user is reminded at a time of his/her own preference for washing hands.


\section{Evaluation}
\label{sec:eval}
We performed a thorough evaluation to demonstrate the performance superiority achieved by \sys~compared to the state-of-the-art smartwatch-based handwashing quality assessment systems. The key questions that drive our evaluations are:

\begin{enumerate}
\item How accurately can \sys~assess whether the user washed hands properly compared to the state-of-the-art systems?

\item In addition to high accuracy, can \sys~minimize the prediction latency of the feedback provided to the user as per the claim of the paper? 

\item How much battery of the smartwatch is consumed by \sys~compared to the state-of-the-art systems?
\end{enumerate}

\subsection{Real-life Dataset Collection}
We collected real-life handwashing data from 14 participants.
The age range of the participants was from 25 to 35 years. Among the 14 participants, 10 were male and 4 were female. 
We developed a data collection app for the Apple watch. The app collects readings from the accelerometer, gyroscope, and magnetometer sensors of the smartwatch with 50Hz sampling rate. 
An instructional video on the WHO handwashing guidelines was shown to the participants so that they could learn the process.
After 3 practice sessions, we collected data of 19 handwashing sessions from each participant.
During each session, the participant wore the smartwatch on the right hand with the data collection app running on it and then performed a handwashing event. The participants were asked to wear the watch in their own ways as well as to perform the handwashing steps in their own ways. All the sessions were video-taped for collecting the ground truth of the handwashing steps.
The collected data were used for the training of \sys.

\begin{figure}[h]
    \centering
    \includegraphics[width=0.8\columnwidth]{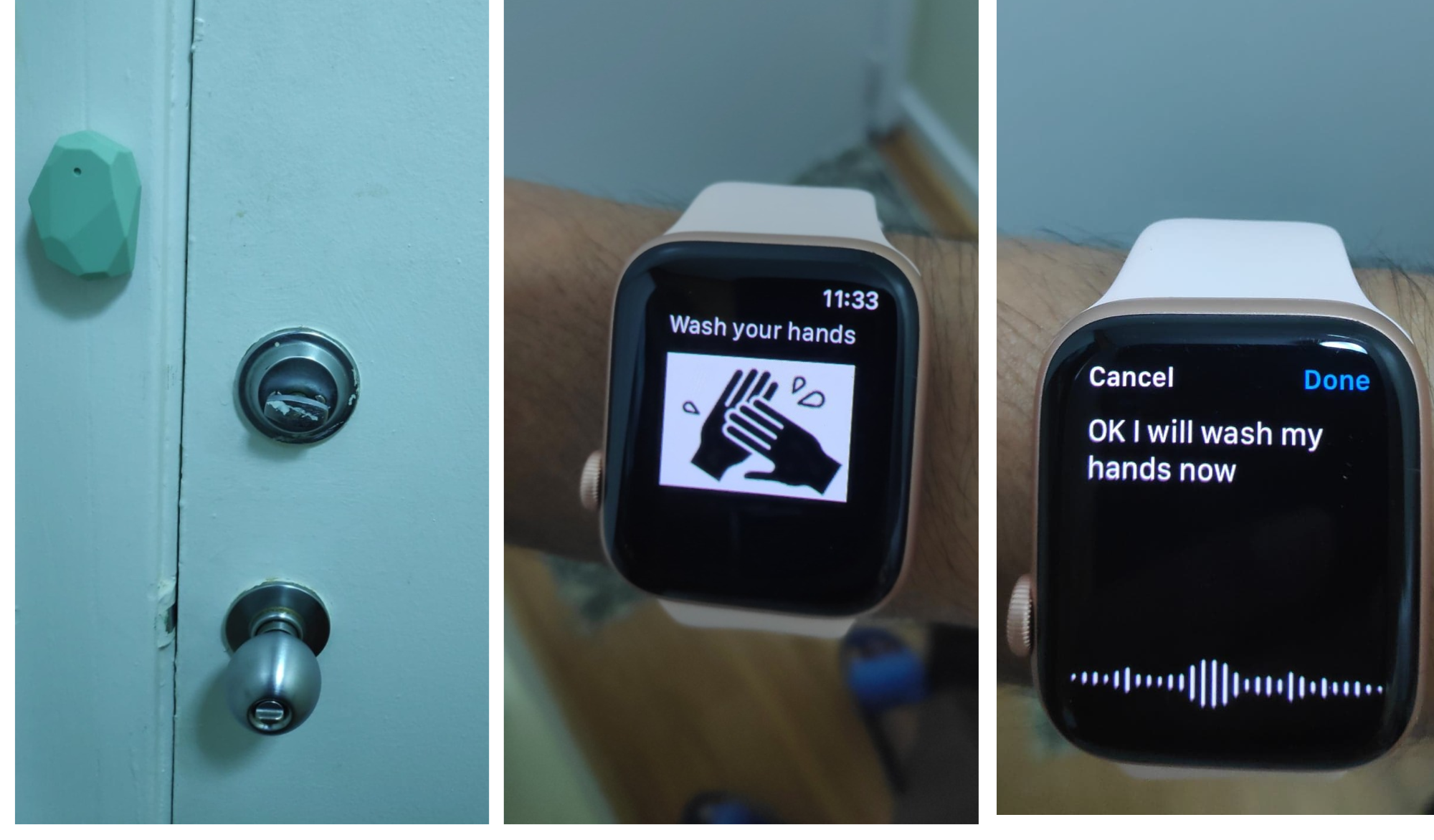}
    \caption{\sys~implementation, with the beacon fitted at the door, user receiving reminder on the smartwatch and responding via voice interaction.}
    \label{fig:iwash_imp}
\end{figure}

\subsection{Training of \sys~}
\label{subsec:train}
We trained the high accuracy model (Section~\ref{subsec:hybrid}) with the collected real-life dataset and then optimized the model for minimizing prediction latency while preserving accuracy (Section~\ref{subsec:opt}). 
The training was performed in a server that has a Intel processor with 10 cores, 256GB RAM, and 4 Nvidia
RTX 2080 Ti GPUs. We used Keras~\cite{chollet2015keras} to implement and optimize the model.

\subsection{Implementation of \sys~}
The implementation of \sys~requires a smartwatch and a proximity Bluetooth beacon. In this work, we implemented \sys~in real-life setting, where we used an Apple watch series 4~\cite{Apple_Watch_Series_4} and an Estimote BLE beacon~\cite{estimote}. The \sys~smartwatch application was developed on the watchOS platform and the trained optimized model (Section~\ref{subsec:train}) was integrated with it. The \sys~application was installed on the Apple watch where it was running on background.

Estimote beacons are one of the most popular, low-energy, proximity bluetooth beacons. An estimote beacon was fitted at the door of the home and the user was wearing the Apple watch. As described in Section 4.5, using the beacon and the smartwatch application, \sys~reminds the user for handwashing and engages the user in a voice interaction (shown in Figure \ref{fig:iwash_imp}). 

We deployed \sys~with the 14 participants for testing. During testing, each participant did 1 session wearing the smartwatch with the \sys~application installed. Following the handwashing, \sys~provided a voice feedback to the user stating if he/she washed hands properly or not. If not, the feedback included if the user missed any steps and if the user did not wash for enough duration for a proper handwashing. 

\subsection{Experimental Setup of \sys~}
\subsubsection{Compared Systems}
We compared the performance of \sys~against several state-of-the-art handwashing quality assessment systems on real-life handwashing data.  These baselines were chosen considering that these systems identified the handwashing steps according to the WHO guidelines. All systems were trained on the collected real-life data (Section 5.1). We evaluated the systems on the test data mentioned in Section 5.3. A short description of the baselines are provided below:

\begin{itemize}
\item WristWash: WristWash \cite{li2018wristwash} is an handwashing quality assessment system that provides a Hidden Markov Model-based analysis for assessing the steps according to the WHO guidelines.

\item H2DTR-NN: This work \cite{galluzzi2015hand} measures the duration and the quality of hand hygiene according to WHO guidelines using different techniques. The best classification results were obtained using a neural network based solution and a k-nearest neighbor based solution. H2DTR-NN represents the neural network based solution presented in the paper.

\item H2DTR-kNN: H2DTR-kNN represents the k-nearest neighbor based solution presented in above paper \cite{galluzzi2015hand}.
\end{itemize}

As per the implementation details described in these papers, both systems require a server where the model for assessing the handwashing quality is loaded and run. When the user performs handwashing, the smartwatch sends the handwashing data to the server. The model running at the server then assesses the handwashing quality, and sends the result back to the smartwatch. In our implementation, the network bandwidth between the smartwatch and the server was 1 Gbps.

\subsubsection{Performance Metrics}

\begin{itemize}
\item Accuracy: To evaluate the performances of the systems in identifying the handwashing steps, we calculate accuracy, a commonly used performance metric. Accuracy of a handwashing step is defined as:
\begin{equation}
\text{Accuracy of a step}= \frac{\text{Number of times the step was identified correctly}}{\text{Total number of times the step appeared in the handwashing data} }
\end{equation}

\item Prediction latency: It measures the average time required by the system to provide the quality result of a handwashing event to the user.

\item Battery usage: It measures the amount of the battery consumed by the handwashing quality assessment system on the smartwatch.
\end{itemize}

\subsection{Experimental Results} 
\subsubsection{Accuracy}
We conducted thorough experiments to evaluate how accurately \sys~assesses the quality of handwashing performed by an user compared to the state-of-the-art systems. Moreover, as previous systems \cite{li2018wristwash} have evaluated the accuracy performance for both user-independent and user-dependent models, we also conducted experiments in both settings to show the performance superiority achieved by \sys.

\noindent \textbf{User-independent Model:}
For the user-independent model, the leave-one-participant-out (LOPO) validation was carried out where the data from 13 participants were used for training, and data from the remaining participant was used for testing. The process was repeated for every participant and the average accuracy score for each of the  handwashing steps was calculated.

\begin{figure}[H]
    \centering
    \includegraphics[width=0.7\columnwidth]{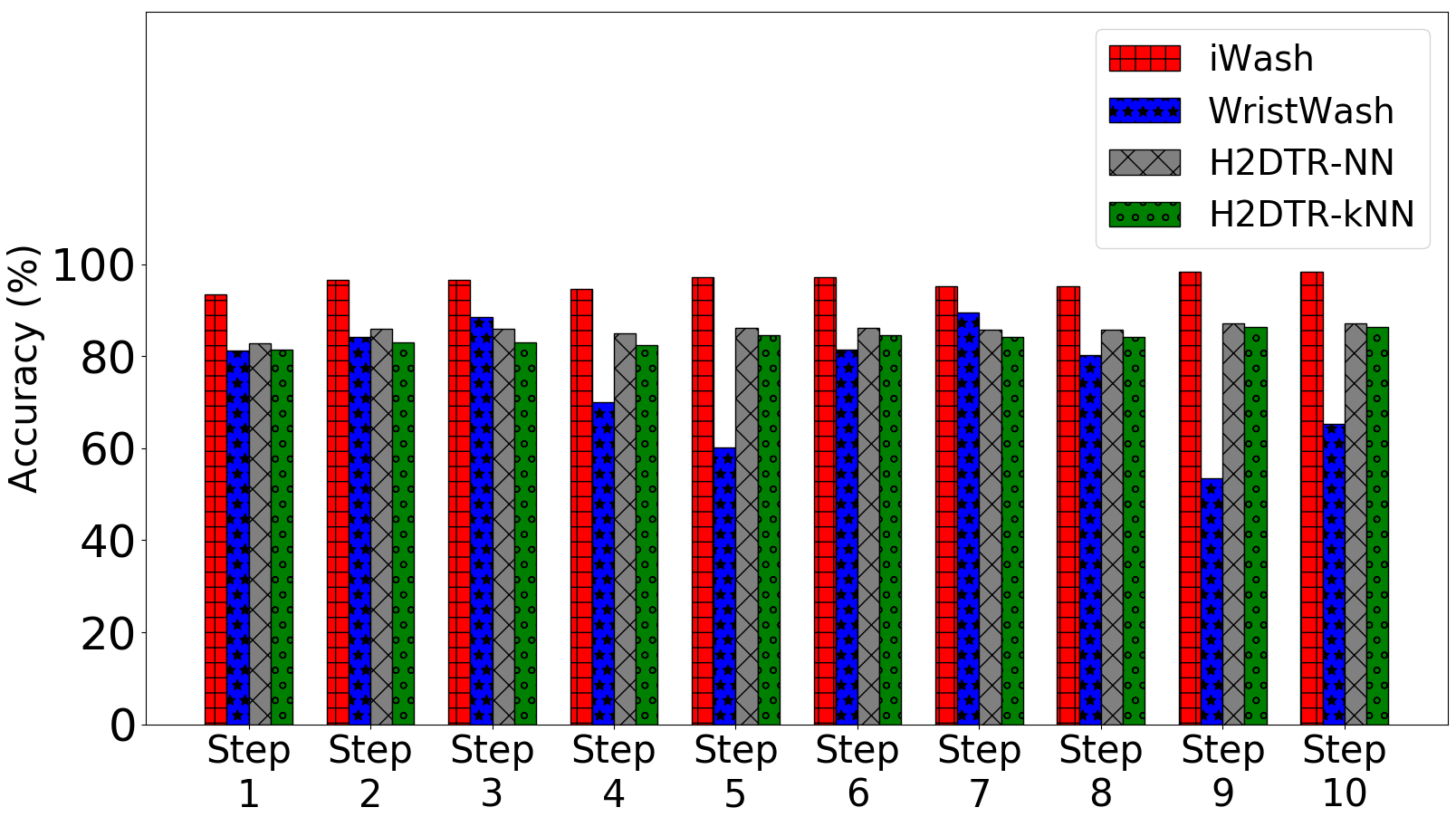}
    \caption{Handwashing steps detection accuracy of the systems for user-independent  model.}
    \label{fig:accuracy_generic}
\end{figure}

Figure~\ref{fig:accuracy_generic} presents the results for the handwashing steps detection accuracy of the systems for user-independent model. On average, \sys~provided around 27.7\%, 12.3\%, and 14.6\% better accuracy than WristWash, H2DTR-NN, and H2DTR-kNN, respectively. This performance improvement occurred because unlike the compared systems, \sys~employs a hybrid deep neural network that effectively captures the correlation present among multiple sensors and the temporal correlation present in sensor data. 

\begin{figure}[h]
    \centering
    \includegraphics[width=0.7\columnwidth]{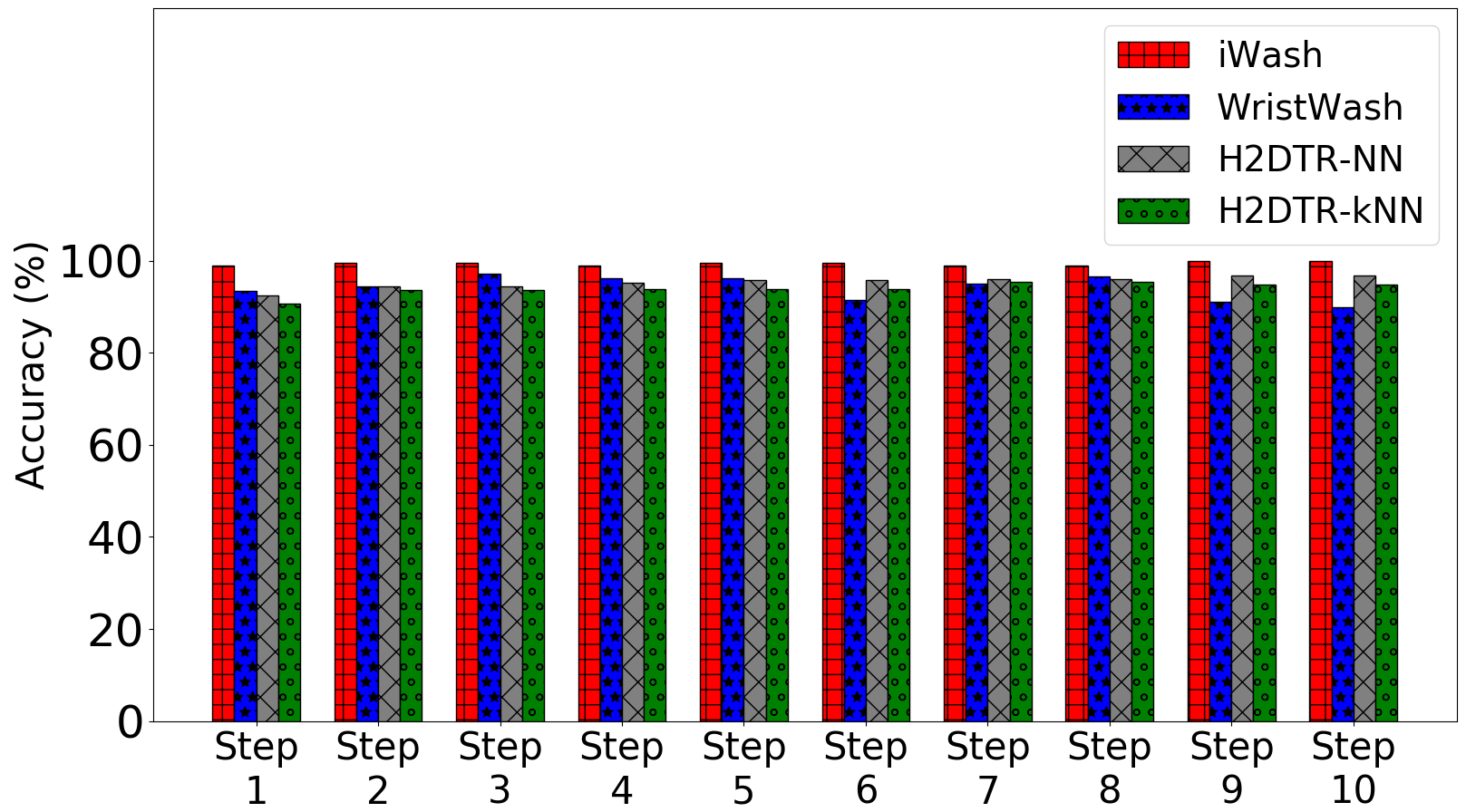}
    \caption{Handwashing steps detection accuracy of the systems for user-dependent model.}
    \label{fig:accuracy_personalized}
\end{figure}

\noindent \textbf{User-dependent Model:}
For the user-dependent model, each of the compared system was trained and evaluated using the data from each participant individually. The process was conducted for every participant and the average accuracy score was calculated for each of handwashing steps.
Figure~\ref{fig:accuracy_personalized} presents the results for the handwashing steps detection accuracy of the systems for user-dependent model. On average, \sys~provided around 15.6\%, 4.2\%, and 5.7\% higher accuracy than WristWash, H2DTR-NN, and H2DTR-kNN, respectively. The superior performance of \sys~occurred for the same reason as described in the user-independent model. Also, each of the systems provided at least 4\% accuracy improvement in the user-dependent model than its corresponding user-independent model. This happened because the user-dependent model that has been trained on only a single individual's handwashing data, can better capture his/her unique features of hand movement during handwashing.

\subsubsection{Prediction Latency}
The time required by the system to 
provide the handwashing quality result to user
was measured for \sys~and for the compared systems.
\begin{figure}[H]
    \centering
    \includegraphics[width=0.7\columnwidth]{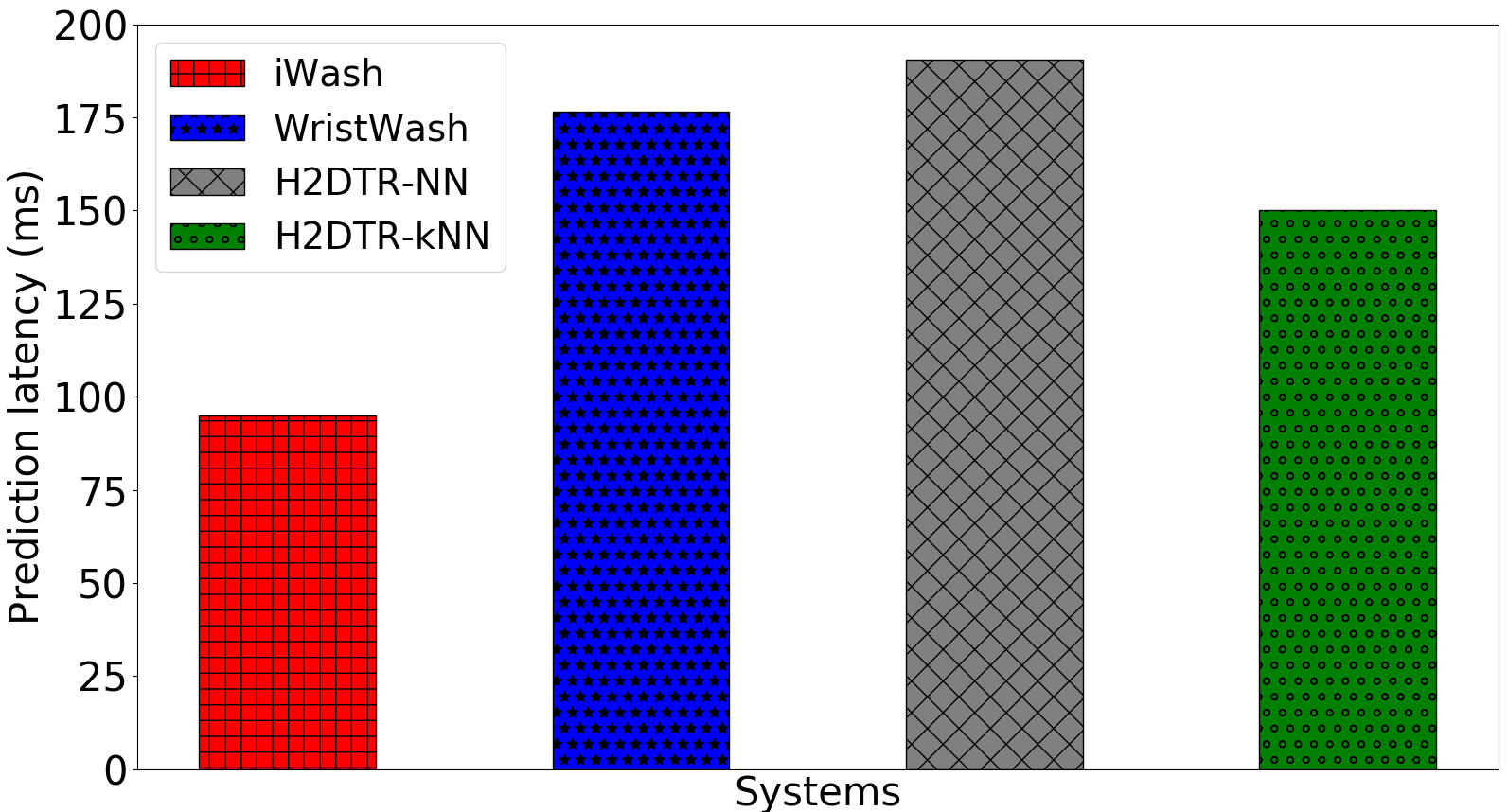}
    \caption{Prediction latency comparison.}
    \label{fig:pred_lat}
\end{figure}

Figure~\ref{fig:pred_lat} shows the performances of \sys~and the compared systems in terms of prediction latency.
\sys~provided around 46.2\%, 50.1\%, and 36.7\% improvement in prediction latency compared to WristWash, H2DTR-NN, and H2DTR-kNN, respectively.
The reason behind this performance improvement is that contrary to the compared systems, \sys~is optimized for both accuracy and prediction latency.

\subsubsection{Battery Usage}
\begin{figure}[h]
    \centering
    \includegraphics[width=0.7\columnwidth]{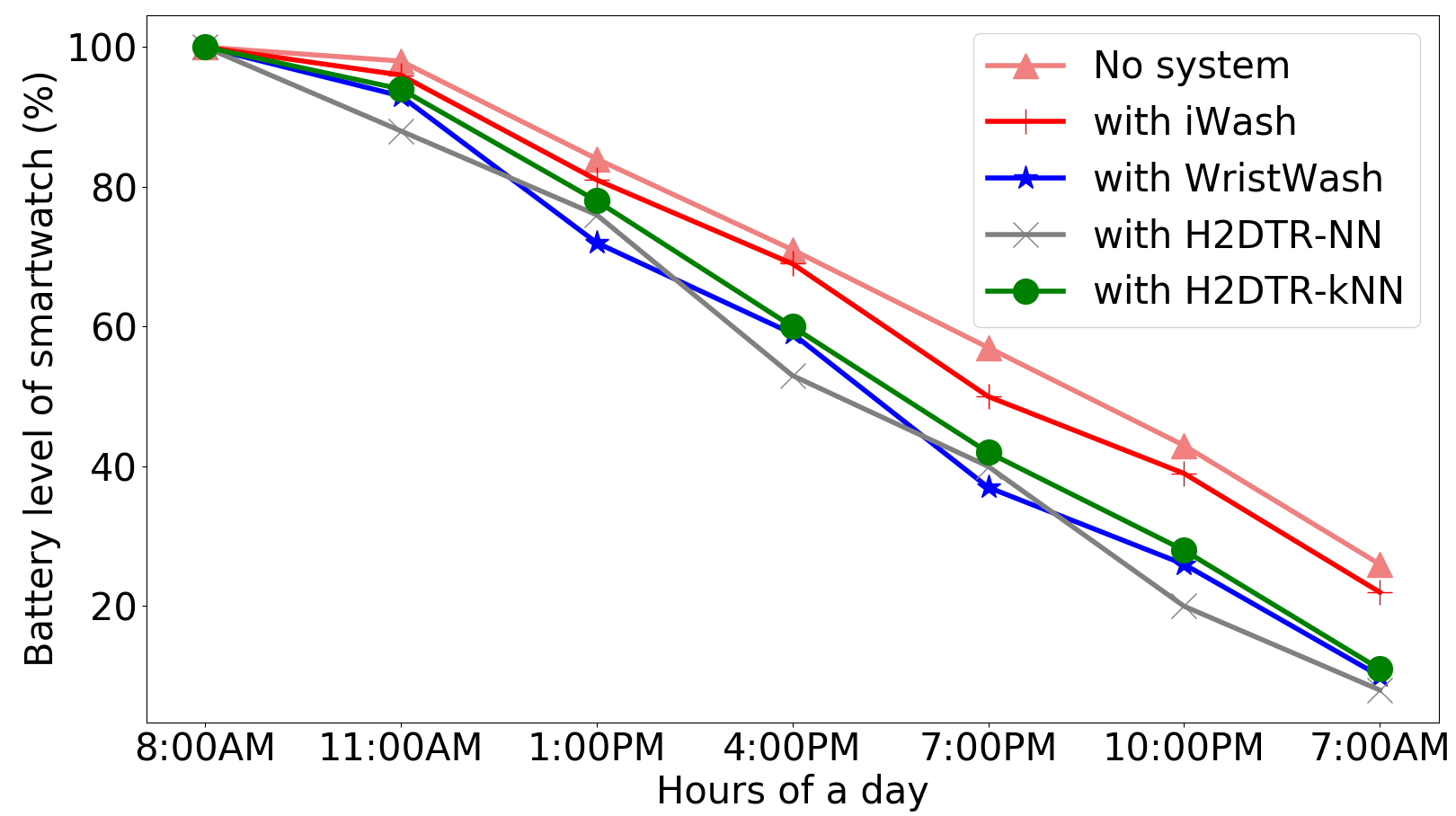}
    \caption{Battery usage comparison.}
    \label{fig:battery}
\end{figure}

To show the effectiveness of \sys~for use in daily life, we evaluated the daily battery usage of the \sys~compared to the state-of-the-art systems, which is demonstrated in  Figure \ref{fig:battery}. Each of compared systems were installed at a time on the smartwatch for evaluation. With each system installed, the battery level of the smartwatch was recorded at various times of the day from 8 AM until 7 AM next morning. Between 7 AM to 8AM the smartwatch was charged every day. The experiment was performed for 7 days on an Apple watch series 4 and we report here the average battery usage. Other than the installed systems, only the default Apple watch applications \cite{appleapps} were running on the watch. A user performed 10 handwashing events on daily basis at various times of the day. The intuition behind this is that, according to previous study, on average a person washes hands 10 times a day \cite{cleaningsumm}.  The result shows that without having any of the systems installed, the battery level of the smartwatch fell to 26\% by the end of the day, whereas using \sys,~the battery level of the smartwatch fell to 22\% by the end of the day. Thus, it is evident that \sys~does not consume any significant battery for the daily life use, while providing a handwashing assessment system to the user on the smartwatch with high accuracy and low prediction latency. On the other hand, using the other compared systems, the battery level dropped to a level of 8\% to 10\% making those systems less effective for daily basis usage. Now-a-days, most of the people use multiple apps on smartwatch throughout a day \cite{jeong2017smartwatch}, this is where the additional battery level for \sys~will be useful. The reason for the performance superiority of \sys~is, unlike the other systems, \sys~is optimized so that it effectively runs on the smartwatch with minimal battery usage.

\subsubsection{Device Adaptivity of \sys}
\label{subsubsec:adaptive}
\begin{figure}[h]
    \centering
    \includegraphics[width=0.7\columnwidth]{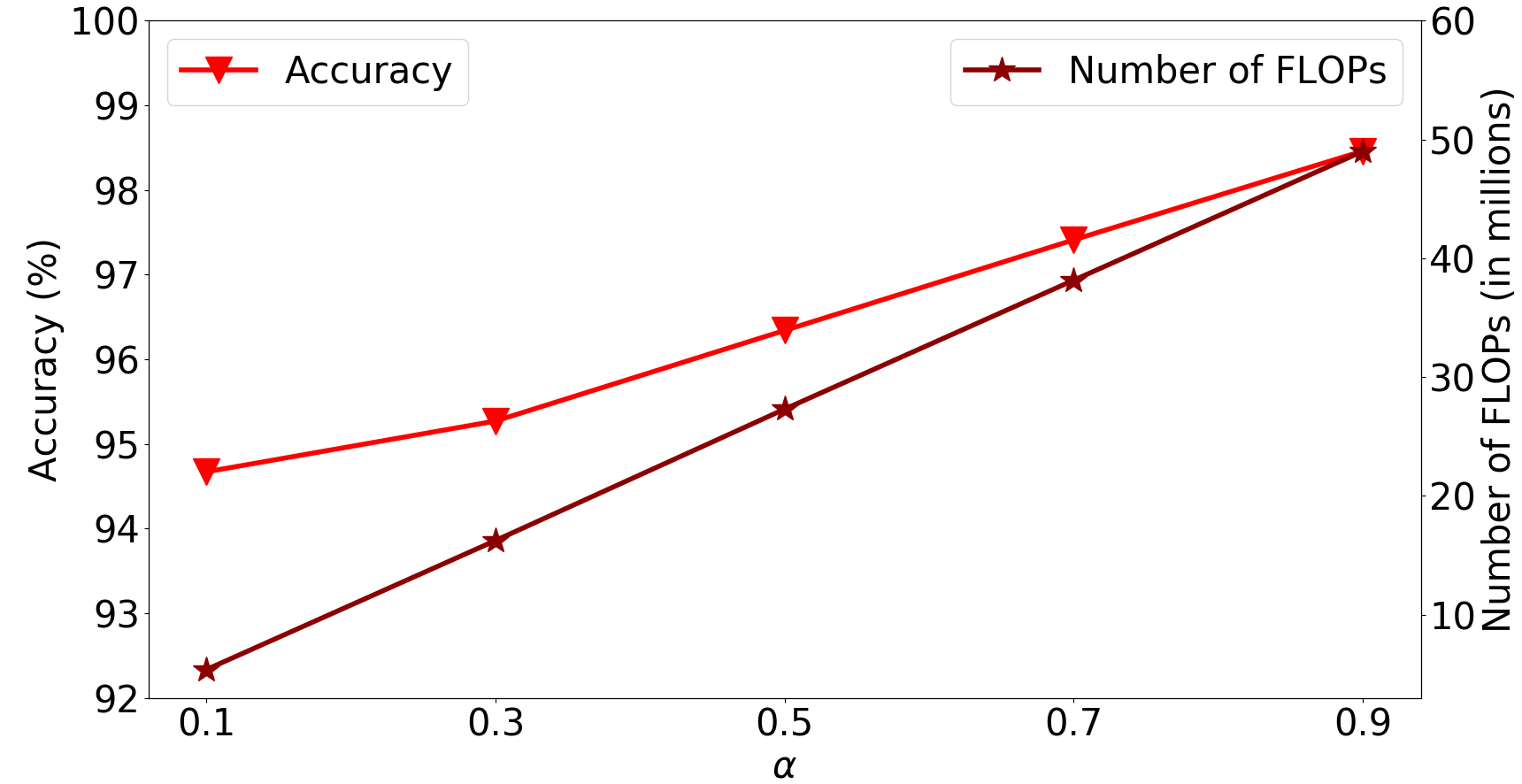}
    \caption{Trade-off between accuracy and number of FLOPs in \sys.}
    \label{fig:device_adaptive}
\end{figure}
The number of FLOPs works as a delegate for prediction latency or consumed energy~\cite{li2020block}. The greater the number of FLOPs, the higher the prediction latency or consumed energy is. 
As explained in Section~\ref{subsec:opt}, the hyper-parameter $\alpha$ introduces a trade-off between accuracy and the number of FLOPs in the deep learning model of \sys. Figure~\ref{fig:device_adaptive} illustrates the trade-off. Here, accuracy refers to the average accuracy of the handwashing steps in the user-independent model. We observe that with the increase of the value of $\alpha$, the accuracy increased, but the number of FLOPs also increased which in turn increased the prediction latency or battery usage of \sys. This occurred because with the increase of $\alpha$, the model was allowed to be more complex facilitating higher accuracy, albeit requiring higher computation. By controlling the value of $\alpha$, we can fulfill the resource constraints of a low-resource device like a smartwatch. In our current implementation, we consider the value of $\alpha$ as 0.5. If a smartwatch has less resources or we need to make further room for other applications running on the device, we can consider the value of $\alpha$ less than 0.5.  From Figure~\ref{fig:device_adaptive}, we observe that, when the value of $\alpha$ decreased from 0.5 to 0.3, number of FLOPs decreased by around 41\%, but accuracy decreased by only 1.2\%. Thus, by sacrificing accuracy a little, we can further decrease the battery usage of \sys. Similarly, for future smartwatches that are expected to have higher battery resources, we can consider a higher value of $\alpha$ that allows achieving higher accuracy. The baseline systems do not have such device adaptivity and, hence, we do not report them in this experiment.

\subsection{Overall Results}
Overall, the results show that \sys~is a comprehensive system that overcomes several limitations of the state-of-the-art handwashing assessment systems in the context of the infectious diseases. Section 5.5.1 shows that \sys~is able to detect the handwashing steps according to the WHO guidelines with a significantly higher accuracy than the other systems for different settings. Also, the results in Section 5.5.2 and 5.5.3 demonstrates that \sys~is designed and implemented in a way optimized for running on a smartwatch, as it provides faster processing time and consumes less battery than the other systems to facilitate real-time feedback provided to the user. Section 5.5.4 shows that \sys~can be tuned to adapt as per the resource availability of the devices unlike the state-of-the-art systems which offers no flexibility.



\section{Related Work}
\label{sec:related-works}

\subsection{Camera-based Handwashing Quality Assessment Systems}
Many past works on handwashing quality assessment are based on a camera or vision based systems. These approaches usually require a video camera or a depth sensor camera to be placed above the sink \cite{llorca2011vision} \cite{xia2015hand} \cite{hoey2010automated} \cite{sure_wash}. The camera records the hand movements performed by the user during the hand hygiene activity. The images or videos from the camera are sent to a cloud server, where the data is processed using vision-based approaches to assess the hand hygiene activities. For example, Llorca et al \cite{llorca2011vision} used such an approach using a RGB camera to access the hand hygiene activities of the healthcare workers. Xia et al \cite{xia2015hand} used similar approaches to recognize hand hygiene poses using RGB-D Videos. However, there are several reasons such approaches are not suitable for daily life settings. First, placing a camera in a bathroom or kitchen is potentially privacy invasive. Second, these approaches are not ubiquitous. For example, a camera-based system will not work when the user is outside the home or when the user follows the hand rubbing approach which does not require a sink. Third, such approaches require substantial installation and maintenance costs \cite{boyce2017electronic}, as well as involves heavy computation for the analysis of images and videos.


\subsection{Wearable-based Handwashing Quality Assessment and Reminder Systems}
Several wearables-based systems \cite{li2018wristwash} \cite{galluzzi2015hand} \cite{wang2020accurate} \cite{mondol2015harmony} have been employed for the assessment of handwashing quality. Such approaches overcome all the limitations of the camera-based systems. It does not require installing any camera on the bathroom or kitchen, and works well even when the user in outside the home. This approach usually collects the wrist movements data during a handwashing event using the wrist-worn sensors and then uses different techniques for detecting the handwashing steps from the event according to the WHO guidelines. Additionally, some works integrate handwashing reminders \cite{mondol2015harmony} on the smartwatch too. WristWash \cite{li2018wristwash} is a wrist-worn platform that provides offline analysis for assessing the steps according the WHO guidelines using Hidden Markov Model-based method. Galluzzi et al \cite{galluzzi2015hand} used different machine learning based techniques for recognizing the handwashing duration and steps according to the WHO guidelines using the wrist-worn sensors. However, there are several reasons why all these systems are not comprehensive enough in the context of infectious diseases. First, most of these works are limited to offline analysis only and none of these provide any feedback to the user whether he/she washed hands properly. Second, none of these systems focused on optimizing the systems for running efficiently on smartwatches in addition to guaranteeing accuracy. Third, none of these systems provide any reminders depending on the user context or offer any interaction in touch-free way. Compared to these systems, we achieve higher accuracy for quality assessment, as well as minimize the smartwatch processing time and battery usage to facilitate the immediate feedback. Moreover, in the context of infectious diseases, our system provide the feedback to the users in real-time if he/she did not wash hands properly, reminds the users when entering home and offers touch-free interaction with the user.




\subsection{Other Approaches}
In the past, adherence to the WHO guidelines was also measured from the direct observation by trained auditors \cite{arias2016assessment} \cite{tschudin2015compliance} \cite{szilagyi2013large}. However, such approaches reported very poor compliance to the WHO guidelines \cite{tschudin2015compliance} \cite{szilagyi2013large}. Moreover, such approaches are not automated and are impractical in the context continuous monitoring and social distancing. Previous works have also used automated dispenser based systems to count the handwashing events electronically \cite{kinsella2007electronic} \cite{marra2010measuring}. However, such approaches are not ubiquitous, and can not detect if the user performed the handwashing properly.



\subsection{Handwashing Applications since the COVID-19 Outbreak}
Since the COVID-19 outbreak, measures for spreading handwashing awareness and to improve handwashing practice have become widespread. In line with that, now-a-days commercially available smartwatches and wearables companies are trying to add handwashing features utilizing the embedded sensors. Samsung has developed a handwashing app to the Galaxy smartwatches \cite{samsung} that will provide reminders to wash hands regularly and track the duration of the handwashing. Google has also added similar features to their smartwatch operating system \cite{google}. Apple recently announced that it will add handwashing features in the next watchOS version, such as detecting handwashing and it's duration automatically \cite{apple}. However, so far none of these companies have found an established solution to assess the quality of handwashing with respect to the standard guidelines or to provide any feedback to the user whether he/she washed hands properly. Moreover, most of these companies are still working on these features, none have brought any comprehensive system so far. Nevertheless, the current trend of the industry shows the promise of smartwatch based handwashing applications for the future. 



\section{Discussion and Future Work}
\label{sec:discuss}

 \begin{itemize}
     \item \textbf{Potential use-cases of \sys:} The performance superiority achieved by \sys~will also be very significant in clinical environments, that is for healthcare workers where hand hygiene measures needs to be strictly followed. A study \cite{stimpfel2019comparison} shows that the daily average work shift duration of an healthcare worker is 12 hours. The battery usage evaluation of \sys~shows that the smartwatch battery level with the \sys~installed and used on a regular basis can last for the whole day. Therefore, \sys~can be effectively worn by a healthcare worker, while it can provide them feedback about their handwashing quality with a high accuracy, which will help reduce the healthcare-associated infections. Moreover, previous studies \cite{jeong2017smartwatch} have shown that the average daily wear times of a smartwatch are typically 8-10 hours depending on the weekend/weekdays. Since \sys~does not consume any significant amount of battery life (Section 5.5.3) compared to not having a handwashing quality assessment system, it will be very effective for improving the hand hygiene practice in daily life.
     
     \item \textbf{Privacy:} The whole system is deployed in smartwatch and requires no connection with any third-party entity (e.g., cloud provider). Thus, handwashing data do not leave the smartwatch and privacy of user data is ensured.
     
     \item \textbf{Evaluation of \sys~for both hands:} Since washing hands involves movement of both hands, using data from both wrists would provide better performance in assessing the quality of the handwashing events. However, it is not practical to wear smartwatches on both hands, particularly in free living context. Our system requires the user to wear the smartwatch only on one wrist. In this work, we evaluated our system on the data collected from the right hand. Considering the fact that our method is agnostic to the placement of the watch, it would also work for the left hand. In future, we will perform evaluations for the left hand too.  
     
     \item \textbf{Larger user study in the context of infectious diseases:} The focus of this work is to develop a comprehensive system for handwashing quality assessment on a smartwatch, that is more accurate, runs faster and consumes less battery compared to the state-of-the-art, as well as integrates context-aware reminder and touch-free interaction with the user. Applying the system on a larger group of people will be the next step to demonstrate its capability in reducing the spread of infectious diseases - which is beyond the scope of this paper.
     
     \item \textbf{Voice-based cognitive assistant:} Modern consumer-grade smartwatches have a built-in speaker and microphone. Using these, previous work \cite{samyoun2019iadhere} used the idea of voice-interactive assistants for stroke patients that can interact with the user about any query regarding the medication and exercise. Similar to that, \sys~can be extended to a voice-based cognitive assistant for any pandemic situation. Besides offering handwashing reminders and voice interaction, in future \sys~can integrate up-to-date public information about the pandemic downloaded to the smartwatch daily, such as, the current status of the pandemic, current restrictions, etc. Users can then, at any time, ask for information from smartwatch on these topics and receive a verbal response.
     
 \end{itemize}

\section{Conclusion}
\label{sec:conclusion}
Proper handwashing is widely acknowledged to be one of the most important activities for reducing the spread of infection. In this paper, we designed and implemented \sys. To the best of our knowledge, \sys~is the most comprehensive system for smartwatch handwashing quality assessment, that achieves significantly higher accuracy compared to the state-of-the-art handwashing quality assessment systems. It runs on a smartwatch with minimal processing time and battery usage facilitating real-time feedback to the user. It also provides routine and context-aware reminders, and offers touch-free interaction. It is also adaptive according to the device requirements without sacrificing accuracy significantly. We collected real-life dataset and extensive evaluations on the dataset demonstrate that \sys~achieves around 12\% higher accuracy, 37\% lower processing time, and 10\% lower battery usage than the state-of-the-art handwashing quality assessment systems.

\bibliographystyle{unsrt}  
\bibliography{references}  


\end{document}